\documentclass[aps,prb,groupedaddress,twocolumn,amsmath,amssymb]{revtex4}
\usepackage{amssymb}
\usepackage{epsfig} 
\usepackage{subfigure} 
\usepackage{graphicx}
\usepackage[english]{babel}
\usepackage{amsmath}
\usepackage{textcomp}
\usepackage{color}
\usepackage{amsfonts}
\usepackage{hyperref}
\usepackage{float}
\usepackage{scalefnt}

\providecommand{\bra}[1]{\color{black}\left<\left.#1\right|\right.\color{black}}
\providecommand{\ket}[1]{\color{black}\left|\left.#1\right>\right.\color{black}}
\newcommand{\etal}{{\em et al.}}
\newcommand{\mb}[1]{\mathbf{#1}}

\begin{document}

\title{Constrained-DFT method for accurate energy level alignment of metal/molecule interfaces}

\author{A. M. Souza,$^1$ I. Rungger,$^1$ C. D. Pemmaraju$^1$\footnote{Present address: Lawrence Berkeley National Laboratory, 
University of California, California, USA.}, U. Schwingenschloegl$^2$ and S. Sanvito$^{1}$}
\address{$^1$School of Physics and CRANN, Trinity College, Dublin 2, Ireland}
\address{$^2$PSE Division, KAUST, Thuwal 23955-6900, Saudi Arabia}

\begin{abstract}
We present a computational scheme for extracting the energy level alignment of a metal/molecule interface, based on constrained 
density functional theory and local exchange and correlation functionals. The method, applied here to benzene on Li(100), allows 
us to evaluate charge transfer energies, as well as the spatial distribution of the image charge induced on the metal surface. We
systematically study the energies for charge transfer from the molecule to the substrate as function of the molecule-substrate 
distance, and investigate the effects arising from image charge confinement and local charge neutrality violation. For benzene 
on Li(100) we find that the image charge plane is located at about 1.8~\AA~above the Li surface, and that our calculated charge 
transfer energies compare perfectly with those obtained with a classical electrostatic model having the image plane located at the 
same position. The methodology outlined here can be applied to study any metal/organic interface in the weak coupling limit at 
the computational cost of a total energy calculation. Most importantly, as the scheme is based on total energies and not on 
correcting the Kohn-Sham quasi-particle spectrum, accurate results can be obtained with local/semi-local exchange and correlation 
functionals. This enables a systematic approach to convergence.
\end{abstract}

\maketitle

\section{Introduction}

Organic/inorganic interfaces are ubiquitous in many different mesoscopic composites of importance in materials science and 
nanotechnology. It is well-known that the performances of organic-based devices, organic or dye sensitized solar cells and 
molecular diodes/transistors only to name a few, depend strongly on the details of the metal/molecule interface \cite{Koch2007}. 
For instance, in organic solar cells the position of the frontier molecular orbitals of the organic light harvesting material with respect 
to the electrode bands is a key design quantity for engineering materials combinations with enhanced light-to-current conversion.
It is then of great importance to have at hand computational tools capable of accurate predictions of levels alignment.  
This is however a tough theoretical problem.

\begin{figure}
\center
\includegraphics[width=0.48\textwidth]{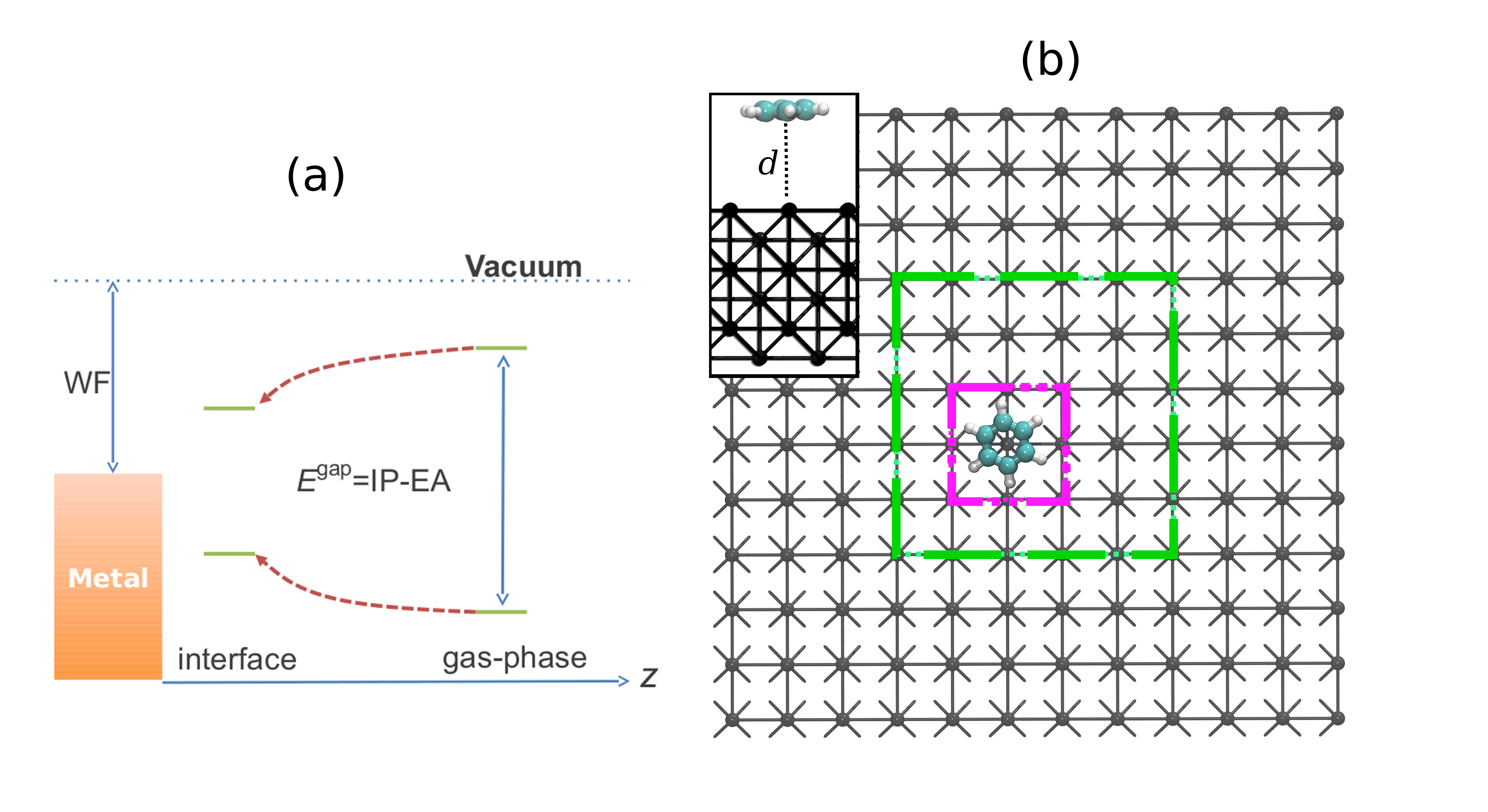}
\caption{(a) Schematic energy level diagram of the frontier orbitals of a molecule approaching a metallic surface. Note the HOMO-LUMO level 
renormalization as a function of the molecule-surface distance, $z$, due to polarization of the metal. (b) Top-view ball-stick representation of a 
benzene molecule at a Li(001) surface for a Li(001) 12$\times$12 supercell. The dashed rectangles show the 3$\times$3 (purple) and 
6$\times$6 (green) supercells. The inset in panel (b) is the side view of the benzene lying flat at a distance $d$ from the surface.}
\label{energy-distance}
\end{figure}

It has been demonstrated experimentally \cite{Search1997,Repp2005,Lu2004,Greiner2011} that the quasi-particle energy gap 
($E^\mathrm{gap}$) of a molecule, defined as the difference between its ionization potential (IP) and electron affinity (EA), gets 
reduced with respect to that of the gas phase by adsorbing the molecule on a polarizable substrate. In a quasi-particle picture the 
IP is the negative of the HOMO energy (HOMO, highest occupied molecular orbital), while the EA corresponds to the energy of 
the LUMO (LUMO, lowest unoccupied molecular orbital). The reduction of the IP and EA of a molecule adsorbed on a metallic 
surface is mainly due to the Coulomb interaction between the added charge on the molecule and the screening electrons 
in the substrate. This interaction leads to a polarization of the surface, so that a surface charge with opposite sign with respect to 
the charge state of the molecule is formed. This non-local feature, called image-charge effect, becomes more relevant as the 
molecule gets closer to the metallic surface. As a consequence the reduction of the IP and the EA, hence of the HOMO-LUMO gap, 
becomes more prominent with the molecule approaching the surface, as schematically illustrated in Fig.~\ref{energy-distance}(a). 

In general, conventional electronic structure theory struggles when predicting the levels alignment at a metal/molecule interface, 
since only rarely non-local correlation effects are explicitly included. This is for instance the case of density functional theory 
(DFT)~\cite{Hohemberg1964,Kohn1965}, today the most widely used method for computing the electronic structure of materials. 
In particular there are two important issues related to DFT and the problem of levels alignment. 
On the one hand, the Kohn-Sham 
eigenvalues cannot be rigorously interpreted as removal energies, Koopman's theorem in general does not apply, so that the 
Kohn-Sham spectrum cannot be taken as a quasi-particle spectrum. The only exception is the HOMO (but not any of the HOMO-$n$
levels), which can be associated to the negative of the IP~\cite{PhysRevB.18.7165,PhysRevLett.49.1691,PhysRevLett.51.1884}. 
Even leaving interpretative issues 
aside in practice the Kohn-Sham energy levels often are not a good representation of the true excitation spectrum of a material,
namely they are not found at the correct energy position. On the other hand, in static DFT the standard approximations to the
exchange and correlation functional, including the local density approximation (LDA)~\cite{Alder1980}, hybrid functionals \cite{b3lyp} 
or explicitly self-interaction corrected ones \cite{Pemmaraju2007,Das2011}, do not include or they do but just poorly, non-local 
correlation effects. This means that, although some of the functionals can predict with satisfactory accuracy the energy levels of the
molecule in the gas phase, they all fail in describing properly the level renormalization as the molecule approaches the surface.
For instance in the LDA there is no change in the HOMO-LUMO gap as a molecule gets closer to a metallic surface~\cite{Neaton2006}.

A conceptually straightforward way to include such non-local correlation effects in the description is that of using many-body
perturbation theory, namely the GW approximation constructed on top of DFT~\cite{Search1973,Hybertsen1986,Onida2002}. 
This approach has been used in the last few years for predicting levels 
alignment~\cite{Neaton2006,Garcia-Lastra2011,Garcia-Lastra2009,Tamblyn2011,Rignanese2001,Strange2012}, in general 
with a good success. The drawback of the GW scheme stays with its computational overheads, which limit the system size
that can be tackled. This is particularly critical for the problem at hand since the typical simulation cells for a molecule on a 
surface are in general rather large. Furthermore, as the image charge may spread well beyond the size of the molecule
investigated, one may even require cells significantly larger than those needed to physically contain the molecule. 

Alternatives to the GW approach, which to some degree also go beyond taking the simple DFT Kohn-Sham spectrum, include 
scissor operators (the DFT+$\Sigma$ approach) \cite{Ferretti2005,Quek2011,Garcia-Suarez2011,Quek2007,Mowbray2008,Abad2008}, 
where the HOMO and LUMO eigenvalues are shifted to match values obtained either from experimental data or from separate total 
energy difference calculations ($\Delta$SCF) plus classical image charge models, and modified $\Delta$SCF schemes 
\cite{Gavnholt2008,Sau2008}. 

Among the various possibilities constrained DFT (CDFT) represents a conceptually different approach to the problem. The idea behind 
CDFT is that one can always define an appropriate density functional, implementing a given desired constraint on the charge
density~\cite{Dederichs1984} (e.g. one can demand that an electron is localized on a particular group in a molecule). This is obtained
by introducing an appropriate external potential in the Kohn-Sham equations. The crucial point is that the approach is fully variational,
meaning that the energy minimum of the constrained functional represents the ground state of the system under that particular 
constraint~\cite{Kaduk2012,Wu2005,Wu2006}. The method allows, for example, to access energies and electron density distributions 
of charge transfer states of a given system, and has been successfully applied to the study of long-range charge transfer excitations 
between molecules \cite{Kaduk2012,Wu2006a,Yeganeh2010}. In the present study we apply CDFT to the investigation of the energy 
level alignment of metal/molecule interfaces. In relation to this problem CDFT has two main advantages. Firstly, since CDFT is based
on total energy differences it does not present the conceptual problems of interpreting the Kohn-Sham eigenvalues as a true 
quasi-particle spectrum. Secondly, one has to note that the total energy, even in the case of local functionals, is a rather accurate
quantity, in contrast to the charge density that local functionals usually tend to over-delocalize. This means that a theory that improves
the charge density but that relies on the total energy is expected to be accurate.
 
The present paper is organized as follows. Firstly we provide a description of the CDFT method used with details on how the constrained 
is imposed. Then results for a specific system consisting of a benzene molecule deposited on a Li(100) surface are presented, focusing 
on charge transfer energies, and hence level alignment. In particular we evaluate the quantitative accuracy of the results as a function of 
the distance between the molecule and the surface, and its dependence on a set of parameters such as the system size and the boundary 
conditions (periodic versus finite). Towards the end we evaluate the changes in the electron density caused by the net charge on the molecule, 
and in particular we determine the position of the image charge plane as function of molecule-surface distance. We then use the calculated 
image charge plane position in a classical model for the energy level shifts and compare our \textit{ab initio} energies to available GW results. 

\section{Method}

In the Kohn-Sham (KS) framework\cite{Kohn1965} the total energy (in atomic units) is given by
\begin{equation}
\begin{split}
 E[\rho]=\sum^{\alpha,\beta}_{\sigma}\sum_i^{N_{\sigma}}\bra{\phi_{i\sigma}}-\frac{1}{2}\nabla^2\ket{\phi_{i\sigma}}+
\\+\int d\bf{r}\, \textit{v}_{\textit{n}}(\bf{r})\rho(\bf{r})+{\textit J}[\rho]+{\textit E}_{\text{xc}}[\rho^{\alpha},\rho^{\beta}]\:,
\label{ks-energy}
\end{split}
\end{equation}
where $J$ is the Hartree energy, $E_{\text{xc}}$ is the exchange-correlation energy, $\textit{v}_n(\bf{r})$ is the external potential, $\rho^\sigma(\bf{r})$ 
is the electronic density for spin $\sigma=\uparrow, \downarrow$ of $N_{\sigma}$ electrons ($\rho=\rho^\uparrow+\rho^\downarrow$) and the set 
$\left\{\ket{\phi_{i\sigma}}\right\}$ contains the KS wavefunctions that minimize the energy. A generic constraint on the charge density is that 
there is a specified number of electrons for each spin, $N_\text{c}^\sigma$, within a certain region of space. This can be written as 
\begin{equation}
 \int w^{\sigma}_{\text{c}}(\bf{r})\rho^{\sigma}(\bf{r})\textit{d}\bf{r}=\textit{N}_\mathrm{c}^\sigma\:,
 \label{constr}
\end{equation}
where $w_\mathrm{c}^\sigma(\bf{r})$ is a weighting function that describes the spatial extension of the constraining region. In the simplest case 
$w_\mathrm{c}^\sigma(\bf{r})$ can be chosen to be equal to 1 within a certain volume and 0 elsewhere. In order to minimize the KS total 
energy of Eq.~\ref{ks-energy} subject to the constraint of Eq.~\ref{constr}, an additional spin-dependent term, proportional to the Lagrange multiplier 
$V_\mathrm{c}^\sigma$, is added to the energy. A new functional is thus defined to be
\begin{equation}
W[\rho,V_\mathrm{c}]=E[\rho]+\sum_{\sigma}V_\mathrm{c}^\sigma\left(\int w^{\sigma}_{\mathrm{c}}(\bf{r})\rho^{\sigma}(\bf{r})\textit{d}\bf{r}-
\textit{N}_\text{c}^\sigma\right)\:.
\label{new-functional}
\end{equation}
\\
When $\rho$ satisfies the constraint in Eq. \ref{constr} then $E[\rho]=W[\rho,V_\mathrm{c}]$ by construction. Up to the $\rho$ independent term 
$\sum_\sigma V_\mathrm{c}^\sigma N_\mathrm{c}^\sigma$, $W[\rho,V_\mathrm{c}]$ is the ground state energy of a system with an additional 
spin-dependent external potential $V_\mathrm{c}^\sigma w_\mathrm{c}^\sigma(\bf{r})$. The KS equations with this extra potential are then given 
by
\begin{equation}
\begin{split}
 \left[-\frac{1}{2}\nabla^2 + \textit{v}_\textit{n}(\bf{r})+\textit{v}_{\text{xc}}^\sigma(\bf{r})+\textit{V}_\text{c}^\sigma\textit{w}_\text{c}^\sigma(\bf{r})+\right.\\\left.+\int \frac{\rho(\bf{r'})}{\left|\bf{r}-\bf{r'}\right|}\textit{d}\bf{r'}\right]\phi_i^\sigma(\bf{r})=\epsilon_i\phi_i^\sigma(\bf{r})\:, 
\label{ks-equations}
\end{split}
\end{equation}
where $\textit{v}_{\text{xc}}^\sigma$ is the exchange and correlation potential. As in standard Kohn-Sham DFT the electron density is 
constructed from the occupied Kohn-Sham eigenvectors, $\left\{\phi_i^\sigma(\bf{r})\right\}$, until self-consistency is achieved.
In this particular case the self-consistency has also to guarantee that the constraint set by Eq.~(\ref{constr}) is satisfied. The minimization 
then proceeds as follows. Firstly, as in the standard Kohn-Sham scheme, an initial charge density is defined and then updated until 
the Kohn-Sham equations are satisfied self-consistently. Secondly, at every self-consistent step in this update of the charge density a 
second self-consistent loop is performed, 
where for a given input density, $\rho(\bf r)$, the value of $V_\mathrm{c}^\sigma$ is updated until the output charge density obtained via 
solution of Eq.~(\ref{ks-equations}) satisfies the constraint of Eq.~(\ref{constr}). This second step is performed following an optimization 
scheme suggested in Ref.~[\onlinecite{Wu2005}]. Updating $V_\mathrm{c}^\sigma$ in this way ensures that at each self-consistent step 
and therefore also at convergence the constraint is fulfilled.

This methodology was implemented in the DFT package {\sc siesta}~\cite{Soler2002}. {\sc siesta} uses a linear combination of atomic orbitals 
(LCAO) basis set, so that, instead of defining the constraining region in real space via the function $w_\mathrm{c}^\sigma(\bf r)$, we define it 
over the LCAO space. This means requiring that the total charge projected onto a given set of basis orbitals is equal to $N_\mathrm{c}^\sigma$. 
For this aim we have implemented both the L\"owdin~\cite{Lowdin1950,Wu2006} and the M\"ulliken~\cite{Mulliken1955} projection schemes.
{A detailed description of the implementation is given in the Appendix.} 
At the quantitative level our results depend somewhat on the projection method employed, and it has been shown that usually the L\"owdin 
scheme gives the most accurate results \cite{Mulliken1955,Kaduk2012}.

Using such a CDFT approach we can evaluate the charge transfer energy between the molecules and the metal surface, and hence the 
position of the frontier energy levels with respect to the Fermi energy of the metal. For a given substrate size and perpendicular distance, $d$, 
between the molecule and the surface atoms, first a standard DFT calculation without constraints is performed. This determines the total 
ground state energy of the combined molecule+substrate system, $E(\mathrm{mol}/\mathrm{sub};d)$, and the amount of charge 
present on each fragment, one fragment being the molecule and the other the substrate. In our calculations we consider $d$ ranging 
from 4\AA \ to 14\AA, where the molecule is only weakly coupled to the substrate, so that the amount of charge on each fragment is a
well defined quantity. Although CDFT is designed for arbitrary geometries and constraints, in the case of overlapping fragments the amount 
of charge localized on each fragment becomes ill defined and the results have to be taken with care~\cite{Kaduk2012}. 

The next step consists in performing a new DFT calculation, where the constraint is set in such a way that one electron is removed from the 
molecule and one electron is added to the substrate. The total energy of such charge transfer state is $E(\mathrm{mol}^{+}/\mathrm{sub}^{-};d)$. 
Hence the charge transfer energy needed to transfer one electron from the molecule to the substrate, $E_\mathrm{CT}^+$, is given by
\begin{equation}
E_\mathrm{CT}^{+}(d)=E(\mathrm{mol}^{+}/\mathrm{sub}^{-};d)-E(\mathrm{mol}/\mathrm{sub};d)\:.
\label{e-homo}
\end{equation} 
In an analogous way we obtain the charge transfer energy gained by moving one electron from the surface to the molecule, 
$E_{\mathrm{CT}}^-$, as
\begin{equation}
E_{\mathrm{CT}}^{-}(d)=E(\mathrm{mol}/\mathrm{sub};d)-E(\mathrm{mol}^{-}/\mathrm{sub}^{+};d)\:,
\label{e-lumo}
\end{equation} 
where $E(\mathrm{mol}^{-}/\mathrm{sub}^{+};d)$ is the CDFT ground state energy of the configuration where one electron is moved from
the metal surface to the molecule. We note that such procedure always deals with globally charge neutral simulation cells, so that no monopole energy 
corrections are necessary under periodic boundary conditions. 
{Moreover for practical calculations the charge-transfer approach can be expected to be more accurate 
than a calculation using non-neutral cells, where the metal is kept neutral but the molecule is charged. 
For such non-neutral calculations the image charge is formed on the metal surface in an analogous 
way to the charge-transfer setup. However, in order for the metal cluster to be charge neutral 
a charge with opposite sign will also form on the surface of the metallic cluster. Given the finite size of the 
cluster this will lead to additional inaccuracies due to the interaction between the image charge
and such spuriously-confined additional surface charge.}

{Within the charge transfer procedure} we can directly determine the energy level alignment at the interface, since 
-$E_{\mathrm{CT}}^{+}$ (-$E_{\mathrm{CT}}^{-}$) corresponds to the energy of the HOMO (LUMO) with respect to the substrate Fermi energy.
In a similar constrained-DFT approach~\cite{Sau2008} Sau and co-workers calculated the charging energy associated to transferring small 
amounts of charge from the substrate to a specific molecular orbital. The charge transfer energy was then obtained by extrapolation 
to integer charge. In order to avoid the use of such 
extrapolation here we always transfer an entire electron between the molecule and the substrate. Since a CDFT calculation has a 
computational cost only marginally more expensive than that of a standard DFT ground-state one (the CPU time increases by about a 
factor two over the entire self-consistent cycle), CDFT allows us the study of large organic molecules on surfaces. This is a prohibitive task 
for many-body-corrected quasi-particle schemes, such as the GW method.

We apply our CDFT method to compute the energy level alignment of a benzene molecule as a function of its distance, $d$, from a Li(100) 
surface. The calculations are performed using norm-conserving relativistic pseudopotentials~\cite{Troullier1991}, and the LDA~\cite{Alder1980} 
for the exchange-correlation potential. The real space grid is set by an equivalent mesh-cutoff of 300 Ry and the charge density and all the 
operators are expanded over a double-$\zeta$ polarized basis set with an energy-shift of 0.03~eV~\cite{Soler2002}. 

The Li metallic surface is modeled by a 6 atomic layer thick slab. The {\it bcc} primitive unit cell lattice constant is set to 3.51~\AA. We consider 
two types of boundary conditions in the plane of the Li substrate surface, namely periodic boundary conditions (PBC) and non-periodic 
boundary conditions (non-PBC). Furthermore, in order to investigate the finite size effects originating from the size of the Li surface, we consider 
three different cell sizes (for both PBC and non-PBC), namely small (3$\times$3 atoms per layer), intermediate (6$\times$6) and large 
(12$\times$12) (see Fig.~\ref{energy-distance}). In the case of non-PBC the real-space box containing the Li slab supercell has dimensions 
55$\times$55$\times$55~\AA$^3$. This is chosen in such a way that even for the 12$\times$12 slab there is at least 15~\AA \ of vacuum 
between the Li slab and the boundaries of the simulation box. By using a cubic box one can apply Madelung corrections in {\sc siesta}. 
These are necessary since the electrostatic potential is calculated by using periodic boundary 
conditions~\cite{Makov1995}. In the case of PBC the in-plane dimensions are set by the Li supercell size and thus are 
10.56$\times$10.56 \ \AA$^2$, 21.09$\times$21.09\ \AA$^2$ and 42.12$\times$42.12\ \AA$^2$, respectively for the 
3$\times$3, 6$\times$6 and 12$\times$12 cell. The cell dimension in the direction perpendicular to the surface plane is the same 
as for the case of non-PBC, namely 55~\AA. 

We use two different boundary conditions for the Li surface in order to investigate the effects arising from the spurious dipole-dipole 
interaction between image supercells. The size of this spurious interactions can be reduced by increasing the size of the unit cell. For
the PBC setup the dimensions in the plane are set by the Li cluster size, while for non-PBC calculations we use a large simulation cell 
which minimizes the dipole-dipole interaction between periodic images. In this way we can disentangle the effects of changing the extension 
of the Li surface in plane from those associated with the size of the simulation box. Furthermore, in the case of non-PBC, edge effects 
may arise and our aim is to find the required cluster and cell size that gives quantitatively accurate charge transfer energies. 

\section{Results}

In order to determine the energy level alignment between the molecule and the surface, we first need to determine the Li workfunction (WF). 
This is calculated by performing a simulation for the Li slab with PBC and no benzene adsorbed and by taking the difference between the 
vacuum potential and the slab Fermi energy. The so obtained value for the Li(001) WF is 2.91~eV. This is in fair agreement with previous
calculations (3.03 eV)~\cite{Kokko1995}, which have also shown that the Li WF can vary by about 0.5~eV depending on the 
crystallographic orientation of the surface. The experimental values reported for polycrystalline Li vary considerably (2.3-3.1~eV), as 
discussed in Ref. \cite{Lang1971} and references therein.  

In the case of non-PBC the Li substrate is essentially a giant molecule and we can calculate the IP and the EA by means of the $\Delta$SCF 
method, where $\mathrm{IP}=E^{(N-1)}-E^{(N)}$ and $\mathrm{EA}= E^{(N)}-E^{(N+1)}$ ($N$ is the number of electrons in the neutral system). 
The results are listed in Tab. \ref{ip-ea} for the different Li cluster sizes. We note that there is a substantial difference between the IP and the EA, 
resulting in a quasi-particle energy gap of the order of 1 eV for the Li clusters. Such a gap arises because of the charge confinement in the finite 
cluster. In this case electron-electron repulsion energy leads to a decrease of the EA and an increase of the IP as compared to the WF calculated
with PBC. If instead of adding a full electron we add/remove a small fractional charge (0.1 of an electron), electron-electron repulsion energy 
becomes negligible, the gap disappears, and we obtain IP=3.1~eV and EA=3.0~eV. Likewise, the gap is reduced for larger clusters, in which the
electron density of the additional electron/hole can delocalize more. Before investigating the combined molecule/Li system we calculate also 
the IP and the EA for the isolated benzene molecule, and our results are shown in Tab. \ref{ip-ea}. We find the energy gap for the
molecule in the gas phase, $E^\mathrm{gap}=\mathrm{IP}-\mathrm{EA}$, to be in good agreement with experiments 
\cite{Chewter1987,Rienstra-Kiracofe2002}, with other works using the $\Delta$SCF \cite{Rienstra-kiracofe2001} approach and with 
GW calculations \cite{Neaton2006,Nguyen2012}. 
\begin{table}
\scalefont{1.0}
\tabcolsep 2pt
\begin{center}
\caption{Ionization potential (IP), electron affinity (EA) and quasi particle gap ($E^\mathrm{gap}$), in eV, for the three Li 
substrates considered and for the benzene molecule in the gas phase compared with experimental data and GW calculations.}
{
\renewcommand{\arraystretch}{1.6}
\begin{tabular}{cccccccc}
\hline
\hline
                     &\multicolumn{3}{c}{Li substrates}         &&&&                                 \\
                     & 3$\times$3 & 6$\times$6 & 12$\times$12    & \multicolumn{4}{c}{benzene gas phase}   \\
\cline{2-4}\cline{6-8}
                    &\multicolumn{3}{c}{$\Delta$SCF}     &             &  $\Delta$SCF  & $GW$                        & Exp.       \\
\cline{2-4}\cline{6-8}

IP                 & 3.46    & 3.44   & 3.55 &     & 9.56           &  9.23$^a$/9.05$^f$/7.9$^e$  & 9.24$^c$      \\
EA                 & 1.57    & 2.18   & 2.63 &     & -1.45          &-0.80$^a$/-1.51$^f$/-2.7$^e$ & -1.14$^d$  \\
$E^\mathrm{gap}$         & 1.89    & 1.26   &  0.92&     & 11.01          &10.51$^b$/10.55$^f$/10.6$^e$ & 10.38     \\ 
\hline
\hline
\end{tabular}\label{ip-ea}
}
\end{center}
$^a$Ref. [\onlinecite{Nguyen2012}];
$^b$Ref. [\onlinecite{Neaton2006}];
$^c$Ref. [\onlinecite{Chewter1987}];
$^d$Ref. [\onlinecite{Rienstra-Kiracofe2002}];
$^e$Ref. [\onlinecite{Garcia-Lastra2009}];
$^f$Ref. [\onlinecite{Samsonidze2011}]
\\
\end{table}

The benzene/Li interface [see Fig. \ref{energy-distance}(b)] consists of a benzene molecule, in its gas phase geometry, positioned parallel
to the Li surface at a distance $d$. We now evaluate the dependence of the various charge transfer energies (positions of the HOMO and LUMO)
on $d$ for all the different Li supercells as well as for both non-PBC and PBC.
\begin{figure}
\center
\includegraphics[width=0.48\textwidth]{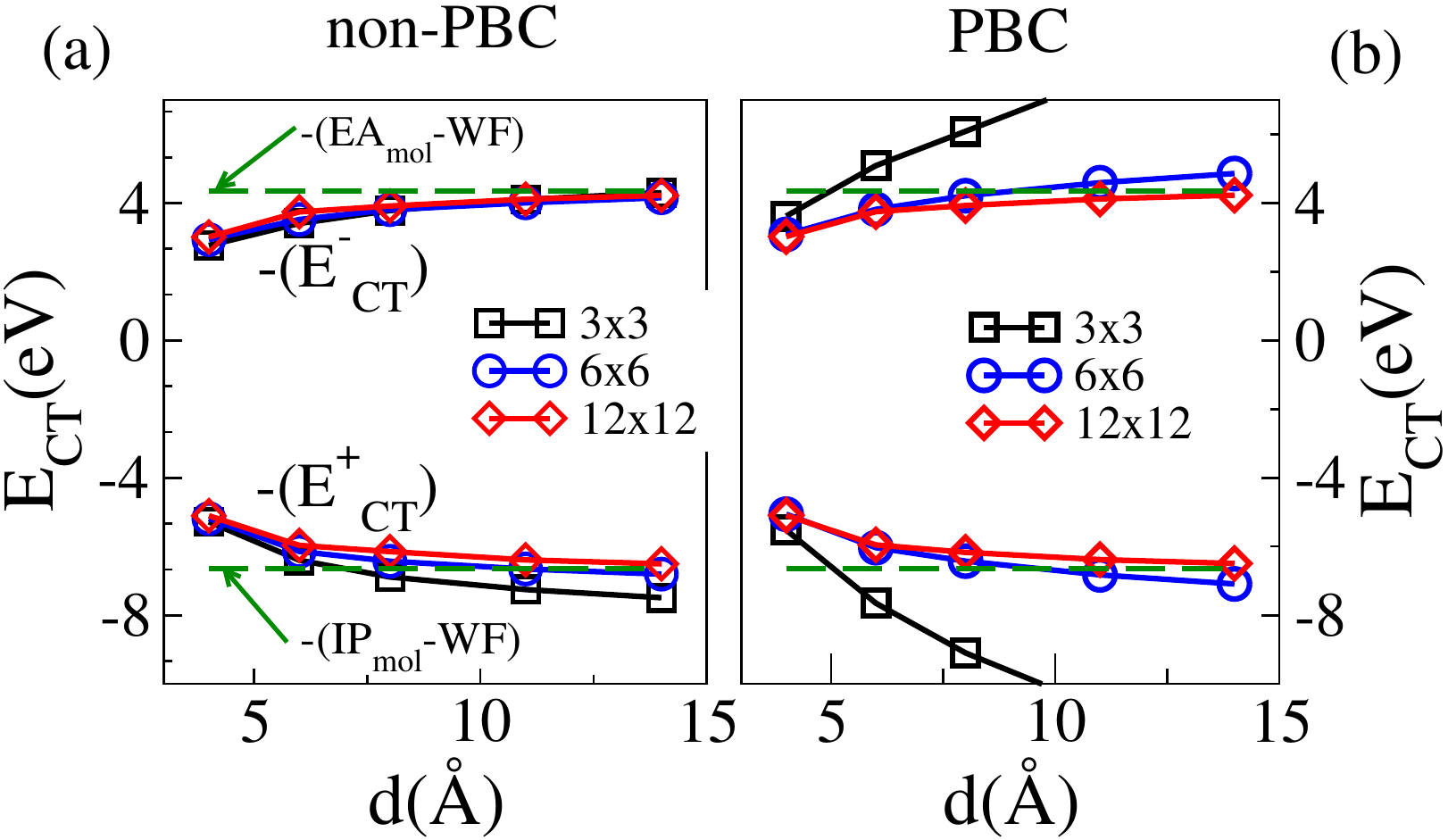}
\caption{Negative of the charge transfer energy, $E^-_{\mathrm{CT}}$, and removal energy, $E^+_{\mathrm{CT}}$, as a function of the 
molecule-surface distance, $d$, for the three clusters considered. Panels (a) and (b) are for non-PBC and PBC calculations, respectively.
The green dashed-lines represent the negative of the $\mathrm{IP}_{\mathrm{mol}}$ and $\mathrm{EA}_{\mathrm{mol}}$ of the isolated 
molecule ($\Delta$SCF calculations) shifted by the calculated Li WF of 2.91~eV.}
\label{gap-distance}
\end{figure}
We start by presenting our results for calculations performed with non-PBC. In Fig.~\ref{gap-distance}(a) we plot $-E_{\mathrm{CT}}^{+}$ and 
$-E_{\mathrm{CT}}^{-}$ as a function of $d$ for all the three Li clusters considered. As expected, due to the electron-hole attraction, the absolute 
value of the charge transfer energy decreases as $d$ gets smaller. This in itself shows that CDFT can capture non-local Coulomb contributions 
to the energy. While for small $d$ the energies of the three different clusters are approximately equal to each other, for large distances they differ
significantly. In order to determine the origin of such deviations we evaluate the same energies in the limit of very large distances ($d\rightarrow\infty$), 
where they become
\begin{equation}
 E_{\mathrm{CT}}^+(\infty)=\mathrm{IP}_{\mathrm{mol}}-\mathrm{EA}_{\mathrm{Li}} \label{ip-ct}
\end{equation}
and
\begin{equation}
 E_{\mathrm{CT}}^-(\infty)=\mathrm{EA}_{\mathrm{mol}}-\mathrm{IP}_{\mathrm{Li}}\:,\label{ea-ct}
\end{equation}
since the interaction energy between the charge on the Li slab and that on the molecule vanishes for $d\rightarrow\infty$. 
The charge transfer energy gap is then given by 
\begin{equation}
\begin{split}
E^{\mathrm{gap}}_{\mathrm{CT}}{(\infty)}&=E_{\mathrm{CT}}^{+}(\infty)-E_{\mathrm{CT}}^{-}(\infty)=\\
&= \mathrm{IP}_{\mathrm{mol}}-\mathrm{EA}_{\mathrm{mol}} + (\mathrm{IP}_{\mathrm{Li}}-\mathrm{EA}_{\mathrm{Li}})\label{gap-ct}\:.
\end{split}
\end{equation} 
While $\mathrm{IP}_{\mathrm{mol}}$ and $\mathrm{EA}_{\mathrm{mol}}$ are independent of the cluster size, this is not the case for 
$\mathrm{IP}_{\mathrm{Li}}$ and $\mathrm{EA}_{\mathrm{Li}}$ (see Table~\ref{ip-ea}). This reflects in the fact that the charge transfer 
energies at large molecule-surface separation varies with the cluster size (see Table~\ref{charge-transfer}).
\begin{table}
\scalefont{1.0}
\tabcolsep 3pt
\begin{center}
\caption{Charge transfer energies (in eV) in the limit of large distances ($d\rightarrow\infty$) for the three molecule/Li cluster cells investigated. 
Values are obtained by evaluating Eqs.~\ref{ip-ct}-\ref{gap-ct} with the $\mathrm{IPs}$ and the $\mathrm{EAs}$ taken from Tab.~\ref{ip-ea}.}
{
\renewcommand{\arraystretch}{1.1}
\begin{tabular}{l c cc}
\hline
\hline
                                   &\multicolumn{3}{c}{Li substrates}   \\
                                   & 3$\times$3       & 6$\times$6          & 12$\times$12     \\
\hline
$E^+_{\mathrm{CT}}{(\infty)}$     & 7.99    & 7.38   & 6.93      \\
$E^-_{\mathrm{CT}}{(\infty)}$     & 4.91    & 4.89   & 5.0      \\
$E^{\mathrm{gap}}_{\mathrm{CT}}(\infty)$     & 12.9    & 12.27   & 11.93      \\ 
\hline
\hline
\end{tabular}\label{charge-transfer}
}
\end{center}
\end{table}

At large distances the variation of $E_{\mathrm{CT}}^+$ with the Li cluster sizes is mainly caused by significant changes in 
$\mathrm{EA}_{\mathrm{Li}}$. Interestingly this is not the case for $E_{\mathrm{CT}}^-$, since $\mathrm{IP}_{\mathrm{Li}}$ is approximately 
the same for all the Li clusters considered. As $d$ gets smaller the extension of the image charge on the Li slab is reduced, so that even small 
clusters are large enough to contain most of the image charge. Therefore the energy differences depend less on the cluster size. For $d$ up to 
about 6~\AA, Fig.~\ref{gap-distance}(a) shows that $E_{\mathrm{CT}}^+$ and $E_{\mathrm{CT}}^-$ are converged even for the small 
3$\times$3 supercell. Since in organic-based devices the first molecular layer deposited on top of the metallic substrate is typically rather close 
to the surface, we expect that in these situations a rather small cluster size will be already sufficient for our CDFT scheme to yield accurately 
converged levels alignment. This means that the CDFT approach is a valuable tool for an accurate evaluation of the electronic structure 
of molecules on surfaces in realistic conditions. Finally, when one looks at larger $d$, it is immediately clear that larger cluster sizes must be
considered. The green dashed lines in Fig.~\ref{gap-distance} correspond to the infinite cluster size limit, for which we have 
$\mathrm{IP}_{\mathrm{Li}}$=$\mathrm{EA}_{\mathrm{Li}}$=WF$\approx$~2.9 eV. It can then be seen that even up to the largest considered 
$d$ of 14 \AA~results obtained for the 12$\times$12 cluster are within the infinite cluster limit (set in the figure by the two green dashed lines), 
so that they can be considered converged.

\begin{figure}
\center
\includegraphics[width=0.45\textwidth]{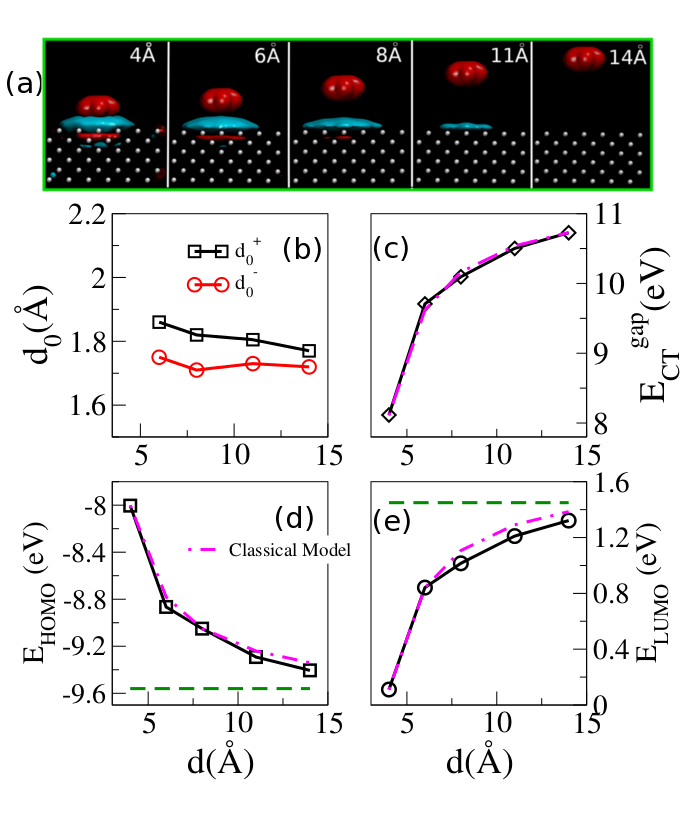}
\caption{Image charge analysis. (a) Isosurface of the difference between the charge densities calculated with DFT (ground state) and CDFT 
(charge transfer state), $\Delta\rho({\bf r})$. Note the formation and the spatial distribution of the image charge. Different panels correspond 
to different molecule/surface distances, $d$. The isosurfaces are taken at 10$^{-4}$e/\AA$^3$. Red isosurfaces denote negative $\Delta\rho({\bf r})$
(electrons depletion), while blue are for positive $\Delta\rho({\bf r})$ (electrons excess). (b) position of the charge image plane taken from the 
surface atoms [see Eq. (\ref{image-charge-plane}] , respectively when one electron,  $d_0^+$, or one hole, $d_0^-$, is transferred from the 
molecule to the Li substrate for the 12$\times$12 PBC calculations as a function of $d$. (c), (d) and (e) are $E_{\mathrm{CT}}^{\mathrm{gap}}$, 
$E_{\mathrm{HOMO}}$ and $E_{\mathrm{LUMO}}$, respectively, as a function of $d$ and compared with the classical model of
Eq.~(\ref{classical-model}). The dashed-green lines are -$\mathrm{IP}_{\mathrm{mol}}$ and -$\mathrm{EA}_{\mathrm{mol}}$ calculated 
with $\Delta$SCF.}
\label{image-charge-iso}
\end{figure}
We now move to the case of PBC, in which there are no edge effects due to the finite size of the cell. Results for the charge transfer 
energies are presented in Fig.~\ref{gap-distance}(b). Although the general trends are analogous to the ones found for the case of
non-PBC, we note that for the 3$\times$3 supercell the changes in the charge transfer energy as a function of $d$ are largely 
overestimated. This is due to the use of PBC, in which the lateral dimensions of the supercell box coincide to those of the Li slab (i.e. 
there is no vacuum). Because of the PBC one effectively simulates a layer of charged molecules and not a single molecule on the surface.
Thus, when the molecules are closely spaced, the charge transfer energy is that of two opposite charged surfaces facing each other (the 
molecular layer and the Li slab). This is significantly larger than that of a single molecule (note that we always compare the charge transfer 
energy per cell, i.e. per molecule). When one increases the size of the supercell and arrives to 12$\times$12, both PBC and non-PBC 
calculations produce the same results. This confirms the observation that the 12$\times$12 supercell is large enough to contain a substantial 
part of the image charge as well as to minimize the Coulomb interaction between repeated supercell images up to $d=$~14~\AA.

From the charge transfer energies we can now obtain an approximate value of the energies of the HOMO and LUMO orbitals, 
by offsetting them with the metal WF, so that $E_{\mathrm{HOMO}}\simeq -(E^+_{\mathrm{CT}}+\mathrm{WF})$ and 
$E_{\mathrm{LUMO}}\simeq -(E^-_{\mathrm{CT}}+\mathrm{WF})$. 
{Note that if the metal substrate is semi-infinite in size, then these relations become exact, since by definition 
the energy required to remove an electron from the metal and that gained by adding it are equal to the 
workfunction. However, in a practical calculation a finite size slab is used, and therefore 
the relations are only approximately valid due to the inaccuracies in the calculated WF for finite systems. As 
shown above, the WF becomes more accurate as the cluster size is increased.}

In the figures \ref{image-charge-iso}(d) and \ref{image-charge-iso}(e) 
we show the calculated values for the 12$\times$12 supercell and PBC obtained by using the Li WF of the infinite slab of 2.91~eV, and
in Fig.~\ref{image-charge-iso}(c) we present $E_\mathrm{CT}^\mathrm{gap}(d)=E_\mathrm{CT}^+(d)-E_\mathrm{CT}^-(d)$.
In order to quantify how the image charge changes the charge transfer energies as a function of $d$, we can write 
$E_\mathrm{CT}^\pm(d)=E_\mathrm{CT}^\pm(\infty)+V(d)$, where the new quantity $V(d)$ corresponds to the energy 
lowering due to the distance dependent electron-hole attraction. It was demonstrated a long time ago~\cite{Lang1973}, by 
using self-consistent DFT calculations, that for flat surfaces $V(d)$ can be accurately approximated by the classical 
image charge energy gain 
\begin{equation}
V(d)=\frac{-q^2}{4(d-d_0)}\:,
\label{classical-model}
\end{equation}
where $q$ is the charge on the molecule, and $d_0$ is the height of the image charge plane with respect to the topmost 
surface atomic layer
\begin{equation}
 d_0(d)=\frac{\int_{d_A}^{d_B} z\ \Delta\rho_{xy}(z;d)\ \mathrm{d}z}{\int_{d_A}^{d_B}\Delta\rho_{xy}(z;d)\ \mathrm{d}z}\:. 
 \label{image-charge-plane}
\end{equation}
In other words $d_0$ can be interpreted as the center of gravity of the screening charge density localized on the metal surface, and in 
general it depends on $d$. Here $\Delta\rho_{xy}(z;d)=\int \mathrm{d}x \mathrm{d}y \Delta\rho({\bf{r}};d)$ and $\Delta\rho({\bf{r}};d)$ is 
the difference between the charge densities of the DFT (ground state) and the CDFT (charge transfer state) solutions for a fixed $d$. 
Note that the charge transfer between the surface and the molecule leads to the formation of a spurious charge layer on the back 
side of the Li slab (i.e. opposite to the surface where the molecule is placed), which is due to the finite number of atomic layers
used to simulate the metal surface. In order not to consider such spurious charge while evaluating the integral in Eq.~\ref{image-charge-plane}, 
the two integration limits, $d_A$ and $d_B$, are chosen in the following way: 1) $d_A$ is taken after the first two $\Delta\rho(d)$
charge oscillations on the back of the cluster, and 2) $d_B$ is the distance at which $\Delta\rho(d)$ changes sign between the top Li 
layer and the molecule (i.e. it is in the vacuum).

Fig. \ref{image-charge-iso}(a) provides a visual representation of the image charge formation as the molecule approaches the surface
and shows isosurface plots of $\Delta\rho({\bf r};d)$ for different distances $d$. Here we present the case in which one electron is removed 
from the molecule and added to the Li surface. As one would expect, the further away the molecule is from the surface the more delocalized 
the image charge becomes~\cite{Lang1973}. Note that the isosurface value is kept constant for all $d$ 
($\Delta\rho({\bf r};d)=10^{-4}$e/\AA$^3$), so that the apparent shrinking of the image charge for $d=$~11~\AA\ simply reflects the fact
that most of the image charge is now spread at an average density smaller than 10$^{-4}$e/\AA$^3$. Likewise, no isosurface contour
appears on the Li slab for $d=$~14~\AA, since now the image charge is rather uniformly spread at low density. In contrast at small $d$
the oscillations of the charge density between the atomic layers of the metallic surface can also be seen. {It can also be seen that at
4 \AA\ the charges on the molecule and the image charge on the Li surface start to overlap. 
Note that for even shorter distances, when the overlap becomes very large, the CDFT approach presented here becomes ill 
defined, since the charge on each fragment is not well defined anymore.}

By evaluating Eq.~\ref{image-charge-plane} we now determine $d_0(d)$ and the results obtained for the 12$\times$12 PBC calculations 
are shown in Fig.~\ref{image-charge-iso}(b) for both electron ($d_0^+$) and hole ($d_0^-$) transfer from the molecule to the surface. 
The average $d_0$ values are 1.81~\AA \ and 1.72~\AA \ for  $d_0^+$ and $d_0^-$, respectively. Although the two values are similar, they
are not identical. This is consistent with the small band-gap of the Li slab, which indicates that holes and electrons behave differently. The 
average values of $d_0^+$ and $d_0^-$ can now be used to evaluate Eq.~\ref{classical-model} for the classical model. The results are 
shown in the figures \ref{image-charge-iso}(c) through \ref{image-charge-iso}(e) and demonstrate that the classical model works remarkably 
well for this system (the calculated slope of both $E_\mathrm{HOMO}(d)$ and $E_\mathrm{HOMO}(d)$ matches almost perfectly that 
obtained by CDFT). It also shows once again that the results for our 12$\times$12 PBC cell are indeed well converged with respect to the 
slab and cell size. 

Finally we make a comparison between our results and those available in the literature for many-body based calculations. We find 
an overall reduction of $E_\mathrm{CT}^\mathrm{gap}$ of 2.5~eV, when the benzene moves from infinity to $d=4.5$~\AA. 
Garcia-Lastra \textit{et al.}\cite{Garcia-Lastra2009} studied the dependence of the frontier quasi-particle energy levels of a 
benzene molecule as a function of the distance to a Li substrate by means of GW calculations. They found an overall reduction 
in $E^\mathrm{gap}$ of $\sim$3.2~eV as compared to the benzene HOMO-LUMO gap in the gas phase, as one can extract from 
Fig.~1(c) of Ref. [\onlinecite{Garcia-Lastra2009}]. The authors also fit their GW results to the classical model, finding the best 
match fitting for $d_0=1.72$~\AA, in very good agreement with our calculated value. There is a small discrepancy in the
results of Ref.~[\onlinecite{Garcia-Lastra2009}], since if one uses the classical model of Eq.~(\ref{classical-model}) with 
$d_0=1.72$~\AA, then the HOMO-LUMO gap reduction should be smaller than 3.2 eV, namely 2.6~eV at $d=4.5$~\AA. Note that the GW
results are obtained for cells much smaller than the converged 12$\times$12 used here. If we now force the classical model to
fit our results for the 3$\times$3 and 6$\times$6 supercells, we will obtain respectively $d_0=2.3$~\AA\ and $d_0=2.1~$\AA,
for a corresponding gap reduction of 3.27~eV and 3.0~eV.
In these two cases however the fit is good at all $d$ only for the 6$\times$6 supercell, while it brakes down for the 3$\times$3
one for $d$ beyond 8~\AA. This is somehow expected since for large molecular coverages (the 3$\times$3 cell) the point-like
classical approximation is no longer valid.

\section{Conclusion}

In summary, we have implemented and applied CDFT to determine the energy levels alignment of metal/organic interfaces in the weak 
electron coupling regime, i.e. for molecules physisorbed on surfaces. In particular we have demonstrated how the frontier energy levels 
of a benzene molecule change, leading to a HOMO-LUMO gap reduction, when the molecule is brought close to a Li(100) surface. This 
effect is due to the screening charge formed on the metal surface. We have then shown that, in order to obtain quantitatively converged 
results, rather large metal cluster sizes are needed for large distances, whereas at small molecule-metal separations smaller clusters 
can also give quantitatively accurate results. Our calculated value for the image charge plane is 1.72 \AA \ and 1.80 \AA \ for EA and IP, 
respectively, in good agreement with the values fitted from GW calculations. Using these distances for the image charge plane height we 
have compared our \textit{ab initio} results with a classical electrostatic model and found good agreement. The approach presented here
offers several advantages over many-body quasi-particles schemes, namely: (i) rather large systems can be calculated, since the computational 
costs are similar to those of standard DFT calculations; (ii) surfaces with arbitrary shapes and reconstruction can be studied, including
defective and contaminated surfaces; (iii) it gives a direct way of determining the position of the image charge for such interfaces. Overall
CDFT applied to the levels alignment problem appears as a promising tool for characterizing theoretically organic/inorganic
interfaces, so that it has a broad appeal in fields such as organic electronics, solar energy devices and spintronics.

\section*{Acknowledgments}

The authors are thankful to the King Abdullah University of Science and Technology (Kingdom of Saudi Arabia) for the financial support through 
the ACRAB project and to the Trinity College High-Performance Computer Center for computational resources.

\section*{APPENDIX: Implementation details}

Our implementation of the constrained DFT approach within SIESTA follows the prescription of Wu \etal ~
described in Ref. \onlinecite{Wu2006b}. Accordingly, we begin by defining a set of constraints on the 
electronic spin density of the form
\begin{equation}\label{cons}
\sum_\sigma \int w_{\mathrm{k}}^\sigma(\mb{r})\rho^\sigma(\mb{r})d\mb{r}-N_{\mathrm{k}}=0\:,
\end{equation}
wherein $\sigma=\uparrow,\downarrow$ represents the spin index, $w_{\mathrm{k}}^\sigma(\mb{r})$ is a weight function 
corresponding to the constraint k, defining the property being constrained and $N_{\mathrm{k}}$ is the constraint value. 
The total electron density is given by
\begin{equation}
\rho(\mb{r})=\sum_\sigma \rho^\sigma(\mb{r})=\sum_\sigma\sum_i^{N_\sigma}|\phi_i^\sigma(\mb{r})|^2\:,
\end{equation}
where $N_\sigma$ is the number of occupied Kohn-Sham orbitals $\phi_i^\sigma(\mb{r})$. A Lagrange multiplier, $V_{\mathrm{k}}$,
is associated to each constraint specified in Eq.~(\ref{cons}). This allows the following modified energy functional to be defined
\begin{equation}
W[\rho,\{V_\mathrm{k}\}]=E[\rho]+\sum_{\mathrm{k}} V_{\mathrm{k}} \bigg[ \sum_\sigma \int w_{\mathrm{k}}^\sigma(\mb{r})\rho^\sigma(\mb{r})d\mb{r}-N_{\mathrm{k}} \bigg]\:,
\end{equation}
with $E[\rho]$ being the standard Kohn-Sham (KS) energy functional given by
\begin{align}\label{KSEnergy}
E[\rho]=\sum_\sigma \sum_i^{N_\sigma} \langle \phi_i^\sigma|-\frac{1}{2}\nabla^2|\phi_i^\sigma\rangle &+ 
\int v_{ext}(\mb{r})\rho(\mb{r}) +\\
&+J[\rho]+E_{\mathrm{xc}}[\rho^\uparrow,\rho^\downarrow]\:.\nonumber
\end{align}
In Eq.(\ref{KSEnergy}) the first term is the kinetic energy, $v_{ext}(\mb{r})$ is the external potential, $J[\rho]$ is the classical 
Coulomb energy and $E_{\mathrm{xc}}[\rho^\uparrow,\rho^\downarrow]$ is the exchange-correlation energy. The variational 
principle yields the stationary condition for the functional $W$ with respect to the normalized orbitals $\phi_i^\sigma$,
which leads to the following modified Kohn-Sham equations
\begin{align}\label{kseq}
\bigg[-\frac{1}{2}\nabla^2 + v_{ext}(\mb{r})+\int d\mb{r'}\frac{\rho(\mb{r'})}{|\mb{r}-\mb{r'}|} + v_{\mathrm{xc}}^\sigma(\mb{r}) +\\
+\sum_{\mathrm{k}} V_{\mathrm{k}} w_{\mathrm{k}}^\sigma(\mb{r})\bigg]\phi_i^\sigma(\mb{r}) = 
\epsilon_i^\sigma \phi_i^\sigma(\mb{r})\:.\nonumber
\end{align}
Thus the constraints enter the effective KS Hamiltonian in the form of an additional external potential 
$\sum_{\mathrm{k}} V_{\mathrm{k}} w_{\mathrm{k}}^\sigma(\mb{r})$. The ground-state of the constrained KS system is 
obtained by solving Eq.~(\ref{kseq}) in conjunction with Eq.~(\ref{cons}). Wu \etal~have shown \cite{Wu2006b} that the 
functional $W$ is concave with respect to the parameters $V_{\mathrm{k}}$ and that by optimizing $W$ through varying 
$\{V_{\mathrm{k}}\}$, one can find the constraint potential that yields the ground-state of the constrained system. In order 
to optimize $W$, we utilize its first derivate with respect to $\{V_{\mathrm{k}}\}$ given by
\begin{align}\label{grad}
\frac{dW}{dV_{\mathrm{k}}} &= \sum_\sigma \sum_i^{N_\sigma} \bigg(\frac{\delta W}{\delta \phi_i^\sigma} \frac{\partial \phi_i^\sigma}{\partial V_{\mathrm{k}}}
+ c.c \bigg) + \frac{\partial W}{\partial V_{\mathrm{k}}} \\
&=\sum_\sigma \int w_{\mathrm{k}}^\sigma(\mb{r})\rho^\sigma(\mb{r})d\mb{r} - N_{\mathrm{k}} \nonumber
\end{align}
where the stationary condition $\frac{\delta W}{\delta \phi_i^\sigma}=0$ implied by Eq.~(\ref{kseq}) is used. Thus we see 
that the derivative $\frac{dW}{dV_{\mathrm{k}}}$ vanishes automatically when Eq.~(\ref{cons}) is satisfied. 

We now outline the implementation of this formalism for the simulation of electron transfer processes within SIESTA. 
In a typical electron transfer problem one has to partition the system into a donor region (D) and an acceptor region (A). Within
SIESTA, this is done by specifying a certain group of atoms as belonging to D and a second group of atoms as belonging to A.
The constrained calculation then involves the transfer of a specified amount of charge from D to A. In order to partition the
continuous electron density in real space between the A and D regions, we choose an appropriate population analysis
scheme, which in turn determines the form of the weight function~$w_{\mathrm{k}}$ in Eq.~(\ref{cons}). The localized numerical 
orbital basis set within SIESTA is particularly suitable for atomic orbital based population analysis schemes such as the ones 
due to Lowdin \cite{Lowdin1950} and Mulliken \cite{Mulliken1955}. We have implemented weight functions corresponding to 
both the Lowdin and Mulliken schemes within SIESTA. For L\"{o}wdin populations, the number of electrons on a group of 
atoms {\em C} is given by
\begin{align}
N_C&=\sum_{\mu \in C}(\mb{S}^{\frac{1}{2}}\mb{D}\mb{S}^{\frac{1}{2}})_{\mu\mu} \\ \nonumber
   &= \sum_{\nu\lambda}D_{\nu\lambda}\sum_{\mu \in C}S^{\frac{1}{2}}_{\lambda\mu}S^{\frac{1}{2}}_{\mu\nu} \\ \nonumber
   &=\mathrm{Tr}(\mb{D}\mb{w}^L_C)\:, \nonumber
\end{align}
where $\mb{D}$ and $\mb{S}$ are the density and overlap matrices respectively and 
$\mb{w}^L_{C\lambda\nu}=\sum_{\mu \in C}S^{\frac{1}{2}}_{\lambda\mu}S^{\frac{1}{2}}_{\mu\nu}$ defines the L\"{o}wdin weight matrix. 
Similarly, with a Mulliken population analysis, the number of electrons on a group of atoms {\em C} is 
\begin{align}
N_C=\sum_{\mu \in C} (\mb{D}\mb(S))_{\mu\mu}=\mathrm{Tr}(\mb(DS)) 
\end{align}
with the corresponding weight matrix given by\\
\[
w^M_{C\mu\nu}=\begin{cases}
               S_{\mu\nu} &\mathrm{if}~\mu \in C~\mathrm{and}~\nu \in C \\  
               \frac{1}{2}S_{\mu\nu} &\mathrm{if}~\mu \in C~\mathrm{or}~\nu \in C \\  
               0 &\mathrm{if}~\mu \ni C~\mathrm{and}~\nu \ni C \\ 
              \end{cases} 
\]
For charge transfer problems, Wu \etal ~recommend a partitioning of the charge density based on the L\"{o}wdin scheme. 

The self consistent field (SCF) procedure for obtaining the constrained DFT ground-state within the current implementation 
consists of an inner and outer loop. The outer loop is similar to a conventional SCF cycle wherein the orbitals obtained by 
solving the KS equations and the associated self-consistent density are updated. The inner loop consists of optimizing the 
$V_{\mathrm{k}}$ multipliers to ensure that the constraint condition given in Eq.~(\ref{cons}) is satisfied at each step of the 
outer loop. By Eq.~(\ref{grad}), this is equivalent to find the extremes of $W$. Since the derivative of $W$ with respect to the 
$V_{\mathrm{k}}$ is readily available from Eq.~(\ref{grad}), we employ a conjugate gradients (CG) optimization procedure 
to ensure that Eq.~(\ref{cons}) is satisfied. Subsequently, the KS equations are solved and the resulting orbitals are used to 
update the KS density and Hamiltonian in the outer loop. We note that Wu~\etal~ also calculate the second derivative 
(Hessian matrix) of $W$ with respect to the $V_{\mathrm{k}}$ parameters and employ the Newton's method to optimize 
$\{V_{\mathrm{k}}\}$. However, the expression for the second derivatives 
$\frac{\partial W}{\partial V_{\mathrm{k}} \partial V_{\mathrm{l}}}$ depends
explicitly on the KS orbitals, whereas the first derivative [Eq.~(\ref{grad})] involves only the density \cite{Wu2006b}. We therefore 
prefer to work with the gradient alone and employ a CG optimization scheme for the $\{V_{\mathrm{k}}\}$. Thus the overall 
SCF procedure consists of the following sequence of steps: (i) Construct the standard KS Hamiltonian $\mb{H}$ for the current guess 
density. (ii) Obtain the constrained KS Hamiltonian $\mb{H}_C=\mb{H}+\sum_{\mathrm{k}} V_{\mathrm{k}} w_{\mathrm{k}}^\sigma(\mb{r})$ by adding the constraint potential $\sum_{\mathrm{k}} V_{\mathrm{k}} w_{\mathrm{k}}^\sigma(\mb{r})$ from
the previous iteration. (iii) Using the Pulay scheme, mix $\mb{H}_c$ with Hamiltonians from previous iterations to obtain 
$\mb{H}'_C$. (iv) By keeping $\mb{H}'_C$ fixed, optimize $\{V_{\mathrm{k}}\}$ so that the constraints in Eq.~(\ref{cons}) are satisfied.
(v) Solve the KS equations for the Hamiltonian combining $\mb{H}'_C$ and the optimized $\{V_{\mathrm{k}}\}$. The new density matrix $\mb{D}$
thus obtained and the optimized $\{V_{\mathrm{k}}\}$ are used in the next iteration. (vi) Repeat steps (i) through (v) until self-consistency 
is achieved.

\end{document}